\documentclass[twocolumn,floats,floatfix,superscriptaddress,aps,pra]{revtex4}
\usepackage{amsfonts}
\usepackage{amssymb}
\usepackage{amsmath}
\usepackage{calc}
\usepackage{graphicx}
\usepackage{bm}
\usepackage{makecell}

\def\be{ \begin{equation} }
\def\ee{ \end{equation} }
\def\bea{ \begin{eqnarray} }
\def\eea{ \end{eqnarray} }
\def\bse{ \begin{subequations} }
\def\ese{ \end{subequations} }
\def\ba{ \begin{array} }
\def\ea{ \end{array} }

\def\i{\,\text{i}}
\def\e{\,\text{e}}
\def\i{i}
\def\e{e}

\def\to{\rightarrow}

\def\d{\text{d}}

\def\U{\mathbf{U}}

\def\H{\mathbf{H}}

\def\A{\mathcal{A}}

\def\i{{\rm{i}}}

\def\phase{\varphi}

\def\ibid{\textit{ibid.~}}

\def\be{ \begin{equation} }
\def\ee{ \end{equation} }
\def\bea{ \begin{eqnarray} }
\def\eea{ \end{eqnarray} }
\def\bse{ \begin{subequations} }
\def\ese{ \end{subequations} }

\def\U{\mathbf{U}}

\def\d{\text{d}}

\def\H{\mathbf{H}}

\def\A{\mathcal{A}}

\def\i{i}
\def\e{e}

\def\to{\rightarrow}

\newcommand{\ket}[1]{\vert #1\rangle}
\begin{document}

\author{Boyan T. Torosov}
\affiliation{Institute of Solid State Physics, Bulgarian Academy of Sciences, 72 Tsarigradsko chauss\'{e}e, 1784 Sofia, Bulgaria}
\author{Nikolay V. Vitanov}
\affiliation{Department of Physics, St Kliment Ohridski University of Sofia, 5 James Bourchier blvd, 1164 Sofia, Bulgaria}

\title{Broadband composite pulses with phase-error correction}
\title{Broadband composite pulses with error-prone phases}
\title{Broadband composite pulses with errant phases}
\title{Composite pulses with errant phases}
% errant, unreliable, faulty, inaccurate

\date{\today}

\begin{abstract}
Composite pulses --- sequences of pulses with well defined relative phases --- are an efficient, robust and flexible technique for coherent control of quantum systems.
Composite sequences can compensate a variety of experimental errors in the driving field (e.g. in the pulse amplitude, duration, detuning, chirp, etc.) or in the quantum system and its environment (e.g. inhomogeneous broadening, stray electric or magnetic fields, unwanted couplings, etc.).
The control parameters are the relative phases between the constituent pulses in the composite sequence, an accurate control over which is required in all composite sequences reported hitherto.
In this paper, we introduce two types of composite pulse sequences which, in addition to error compensation in the basic experimental parameters, compensate systematic errors in the composite phases.
In the first type of such composite sequences, which compensate pulse area errors, relative phase errors of up to 10\% can be tolerated with reasonably short sequences while maintaining the fidelity above the 99.99\% quantum computing benchmark.
In the second type of composite sequences, which compensate pulse area and detuning errors, relative phase errors of up to 5\% can be compensated.
\end{abstract}

\maketitle

%%%%%%%%%%%%%%%%%%%%%%%%%%%%%%%%%%%%%%%%%%%%%%%%%%%%%%%%%%%%%%%%%%%%%%%%%%%%%%%%%%%%%%%%%%%%%%%%%%%%%%%%%%%%%%%%%%%%%%%%%%%%%%%%%%%%%%%%%
%%%%%%%%%%%%%%%%%%%%%%%%%%%%%%%%%%%%%%%%%%%%%%%%%%%%%%%%%%%%%%%%%%%%%%%%%%%%%%%%%%%%%%%%%%%%%%%%%%%%%%%%%%%%%%%%%%%%%%%%%%%%%%%%%%%%%%%%%
%%%%%%%%%%%%%%%%%%%%%%%%%%%%%%%%%%%%%%%%%%%%%%%%%%%%%%%%%%%%%%%%%%%%%%%%%%%%%%%%%%%%%%%%%%%%%%%%%%%%%%%%%%%%%%%%%%%%%%%%%%%%%%%%%%%%%%%%%
\section{Introduction\label{Sec:intro}}

Since their invention in nuclear magnetic resonance (NMR) 40 years ago  \cite{Levitt1979, Freeman1980, Levitt1982, Levitt1983, Tycko1984, Tycko1985, Shaka1984, Levitt1986, Wimperis1994} composite pulses have established themselves as one of the most accurate, robust and flexible technique for coherent control of quantum systems.
In recent years, they have enjoyed steadily growing interest in quantum information \cite{ions, Ivanov2011, Ivanov2015} and quantum optics \cite{Torosov2011PRA,Torosov2011PRL,Schraft2013,Genov2014PRL}.
Curiously, the concept of composite sequences has been well known in polarization optics since the 1940s \cite{West1949,Destriau1949,Pancharatnam1955, Harris1964,McIntyre1968,Ivanov2012,Peters2012,Dimova2013,Dimova2014},
where achromatic wave plates or polarization filters can be constructed by a set of ordinary wave plates with their fast (or slow) polarization axes rotated at specific angles with respect to each other.
Recently, the composite idea has been extended also to frequency conversion processes in nonlinear optics \cite{Genov2014jopt,Rangelov2014}.

Composite pulses offer a unique combination of ultrahigh accuracy, well below the error threshold (often referred as $10^{-4}$) in quantum computation, with robustness to parameter errors similar to adiabatic passage techniques \cite{Vitanov2001ARPC}.
Moreover, they offer great flexibility in shaping the excitation profile, or even the propagator, in essentially any desired manner --- a feature, which is not available in any other quantum control method.

%Although the required total composite pulse area is typically a few times larger than the one used by the resonance techniques, it is still significantly less than the typical pulse areas in the adiabatic techniques.

%%
%Composite pulse sequences have been invented in nuclear magnetic resonance (NMR) nearly 50 years ago \cite{Levitt1979, Freeman1980, Levitt1982, Levitt1983, Tycko1984, Tycko1985, Shaka1984, Levitt1986, Wimperis1994}.
%In fact, the idea of composite sequences has been known in polarization optics since the 1940's.
%By stacking several ordinary wave plates of the same or different material one can design either achromatic (broadband) polarization retarders or polarization filters,
% by rotating the plates at specific angles with respect to their fast polarization axes \cite{West1949,Destriau1949,Pancharatnam1955, Harris1964,McIntyre1968,Ivanov2012,Peters2012}.
%%

The composite pulse sequence is a finite train of pulses with well-defined phases, which are used as control parameters in order to compensate experimental errors or to shape the excitation profile in a desired manner.
The most ubiquitous composite sequences are the broadband $\pi$ pulses, which produce unit transition probability not only for a pulse area $\A=\pi$ and zero detuning $\Delta=0$, as a single resonant $\pi$ pulse, but also in some (broad) ranges around these values \cite{Freeman1980,Levitt1982,Levitt1983,Levitt1986,Wimperis1994,Torosov2011PRA,Torosov2011PRL,Torosov2018}.
Hence a composite $\pi$ pulse can compensate the pulse area and detuning errors of a single $\pi$ pulse and make a sequence of imperfect pulses look like an ideal $\pi$ pulse.
Among the broadband composite $\pi$ pulses, we note those, which compensate pulse area errors, detuning errors, and both pulse area and detuning errors.
Recently, composite pulses, which compensate experimental errors in any experimental parameter --- universal composite pulses --- have been introduced and experimentally demonstrated \cite{Genov2014PRL}.
Composite $\theta$ pulses, which produce controlled partial excitation with probability $\sin^2(\theta/2)$, are also available \cite{Levitt1986,Wimperis1994,Torosov2019} and they have important applications in quantum computation.

There are two other types of composite sequences: narrowband \cite{Tycko1984,Tycko1985,Shaka1984,Wimperis1994,Torosov2011PRA,Vitanov2011} and passband \cite{Wimperis1994,Kyoseva2013} composite pulses.
Narrowband pulses squeeze the excitation profile inside a narrow range around a certain point in the parameter space and suppress excitation outside of it.
Passband pulses combine the features of broadband and narrowband pulses: highly-efficient excitation in a certain parameter range and very low excitation outside.

In all known composite sequences an accurate control of the composite phases has been presumed.
Because the underlying physical mechanism of composite pulses is constructive or destructive interference of probability amplitudes, the excitation profile is very sensitive to phase errors.
Typically, the relative phase errors must not exceed 0.1-1\% for short composite sequences, and even values less than 0.1\% for long sequences.
This requirement restricts the application of composite pulses to physical platforms wherein such accuracy is possible.
For radiofrequency and microwave driving fields the composite phases are produced by the respective generator and this requirement is usually fulfilled.
In the optical domain such accuracy may become a challenge, especially if the phase shifts are produced by off-resonant electric or magnetic pulses via Stark and Zeeman shifts.

In the present paper, we introduce different composite $\pi$ pulses, which are robust to systematic phase errors of the order of 5-10\%.
We present two types of such composite sequences: (i) sequences that deliver double compensation of simultaneous errors  in the pulse area and the composite phases, and (ii) sequences which produce triple compensation of simultaneous errors in the pulse area, the detuning and the composite phases.

This paper is organized as follows.
In Sec.~\ref{Sec:method} we discuss the mathematical details of the derivation of these new composite pulses.
In Sec.~\ref{Sec:double} the double-compensation sequences are introduced, and the triple-compensation sequences are presented in Sec.~\ref{Sec:triple}.
Finally, Sec.~\ref{Sec:conclusion} wraps up the conclusions.

%%%%%%%%%%%%%%%%%%%%%%%%%%%%%%%%%%%%%%%%%%%%%%%%%%%%%%%%%%%%%%%%%%%%%%%%%%%%%%%%%%%
%%%%%%%%%%%%%%%%%%%%%%%%%%%%%%%%%%%%%%%%%%%%%%%%%%%%%%%%%%%%%%%%%%%%%%%%%%%%%%%%%%%
%%%%%%%%%%%%%%%%%%%%%%%%%%%%%%%%%%%%%%%%%%%%%%%%%%%%%%%%%%%%%%%%%%%%%%%%%%%%%%%%%%%

\section{Description of the method for correction of phase errors}\label{Sec:method}

Here we describe the method for construction of composite pulses that produce excitation profiles which are robust against simultaneous errors in the pulse area and the composite phases (double compensation).
Triple compensation is derived similarly and the specifics are elaborated in Sec.~\ref{Sec:triple}.

The propagator of a coherently driven two-state quantum system, described by the Hamiltonian $\H (t) = \frac12 \hbar [\Omega(t) \sigma_x + \Delta(t) \sigma_z]$, is given by the SU(2) matrix
\be\label{SU(2)}
\U_0 = \left[ \begin{array}{cc} a & b \\ -b^{\ast} & a^{\ast} \end{array}\right],
\ee
where $a$ and $b$ are  the (complex) Cayley-Klein parameters obeying $|a|^2+|b|^2=1$.
For exact resonance ($\Delta=0$), we have $a=\cos(\A/2) $, $b=-\i\sin(\A/2)$, where $\A$ is the temporal pulse area $\A=\int_{t_i}^{t_f}\Omega(t)\d t$.
For a system starting in state $\ket{1}$, the single-pulse transition probability is $p = |b|^2=\sin^2 (\A/2)$.

A phase shift $\phi$ imposed on the driving field, $\Omega(t)\to\Omega(t)\e^{\i\phi}$, is imprinted onto the propagator as
\be\label{U phase}
\U_\phi = \left[ \begin{array}{cc} a & b \e^{\i\phi} \\ -b^{\ast}\e^{-\i\phi} & a^{\ast} \end{array}\right].
\ee
Consider a train of $N$ pulses, each with area $A_k $ and phase $\phi_k$,
\be\label{train}
(A_1)_{\phi_1} (A_2)_{\phi_2} (A_3)_{\phi_3} \cdots (A_N)_{\phi_{N}}.
%(\A_1)_{\phase_1} (\A_2)_{\phase_2} (\A_3)_{\phase_3} \cdots (\A_N)_{\phase_{N}}.
\ee
In the presence of pulse area errors, we have to replace the nominal pulse areas $A_k$ by the actual pulse areas $\A_k = A_k (1+\alpha)$ $(k=1,2,\ldots,N)$, where $\alpha$ is the relative pulse area error.
In the presence of phase errors, each nominal phase $\phi_k$ should be replaced by the actual phase $\phase_k = \phi_k (1+\epsilon)$ $(k=1,2,\ldots,N)$, where $\epsilon$ describes the (systematic) phase errors, the compensation of which is our primary concern here.
In the presence of pulse area and phase errors, the pulse sequence \eqref{train} produces the propagator
\be\label{U^N}
\U^{(N)} = \U_{\phase_{N}}(\A_N) \cdots \U_{\phase_{3}}(\A_3) \U_{\phase_{2}}(\A_2) \U_{\phase_{1}}(\A_1).
\ee
Yet, for the sake of brevity, in the notation of the composite pulse sequence \eqref{train} we shall use the nominal pulse areas $A_k$ and the nominal phases $\phi_k$.

In this paper, based on numerical evidence, we consider composite sequences of an odd number $N=2n+1$ $(n=1,2,\ldots)$  of identical pulses, and nominal pulse area $A_k=\pi$ $(k=1,2,\ldots,N)$.
We also consider symmetric phases, $\phi_k=\phi_{N+1-k}$  $(k=1,2,\ldots,n)$.
Using the invariance of the transition probability to the addition of the same phase shift to all phases (see Appendix \ref{Sec:appendix}), we set $\phi_1= \phi_N=0$.
Hence, the phase-error correcting composite sequences are
\be\label{anagram}
\Phi N =\pi_{0} \pi_{\phi_2} \cdots \pi_{\phi_{n}} \pi_{\phi_{n+1}} \pi_{\phi_{n}} \cdots \pi_{\phi_2}\pi_0,
\ee
and the total propagator turns into
\be\label{U^N2}
\U^{(N)} = \U_{0}(\A)\U_{\phase_{2}}(\A) \cdots \U_{\phase_{n+1}}(\A) \cdots \U_{\phase_{2}}(\A) \U_{0}(\A),
\ee
with $\A = \pi (1+\alpha)$ and %the constituent propagators %$\U_\phase(\pi)$
\be\label{propPhi}
\U_\phase(\A) = \left[ \begin{array}{cc} \cos(\A/2) & -\i\sin(\A/2) \e^{\i\phase} \\  -\i\sin(\A/2) \e^{-\i\varphi} & \cos(\A/2) \end{array}\right].
\ee
%where $\A = A (1+\alpha)$, with $A$ being the nominal (errorless) pulse area and $\alpha$ is the relative pulse area error.

We calculate the product in Eq.~\eqref{U^N2} and expand $U^{(N)}_{11}$ vs $\alpha$ and $\epsilon$ at ($\alpha=0,\epsilon=0)$.
Then we set to zero as many terms as possible in order to obtain a robust excitation profile.
If we denote the $(j,k)$-th multivariate coefficient in the power series as
\be
s_{jk} = \frac{\alpha^j \epsilon^k}{j!k!} \left(\frac{\partial^{j+k}U^{(N)}_{11}}{\partial{\alpha^j}\partial{\epsilon^k}}\right)(0,0),
\ee
one can easily verify that, because of the chosen symmetry in the phases and pulse areas of the composite sequence, we have
\be\label{multivariate}
s_{jk} \equiv 0 \quad (\text{for all even $j$}).
\ee
Hence the first nonzero derivatives with respect to the pulse area error $\alpha$ are the first derivatives ($j=1$), then the third derivatives ($j=3$), etc.
With respect to the phase error $\epsilon$ all derivatives are generally nonzero.
Because the primary objective of the composite pulses is the compensation of pulse area (and detuning) errors, we limit ourselves to the cancellation of the low-order (up to first or second) derivatives with respect to the phase error $\epsilon$, which already provides significant improvement over the existing composite pulses.
%Therefore, in the derivation of the composite phases we first set $s_{10} = 0$, then $s_{11} = 0$, then $s_{30} = 0$, then $s_{31}$, etc. until we run out of phases.
Generally, for $N=2n+1$ pulses we have $n$ different phases, with which we can nullify $n$ different derivatives.

\section{Compensation of pulse area and phase errors}\label{Sec:double}

As a performance benchmark we use the broadband composite pulses of Ref.~\cite{Torosov2011PRA}, which compensate pulse area errors.
They contain an odd number of pulses $N=2n+1$ and symmetric (anagram) phases, which are given a simple analytic formula for arbitrary $N$.
Below we present the new composite pulses which, in addition to pulse area errors, compensate phase errors. 

\subsection{Composite sequences of three pulses}

We first consider a sequence of three pulses.
We calculate the product in Eq.~\eqref{U^N2} and expand $U^{(3)}_{11}$ at $\alpha=0$ and $\epsilon=0$.
We obtain for the first non-zero coefficient in the expansion	
\be
s_{10}=\alpha\frac{\pi}{2}\left[ 1+2\cos(\phi_2)\right] .
\ee
We can nullify this coefficient by setting $\phi_2=2\pi/3$.
The resulting composite sequence $\pi_0 \pi_{\frac23\pi} \pi_0$ coincides with the well-known broadband CP B3 of Eq.~\eqref{B3} \cite{Freeman1980,Torosov2011PRA}. 
Hence we do not obtain additional compensation in the phase error because of the absence of free phases.
(Letting $\phi_3$ be nonzero does not help annul $s_{11}$.)
Longer sequences with $N > 3$, studied below, do allow for such compensation.

\subsection{Composite sequences of five pulses}

%=================================================================
\begin{figure}[tb]
	\includegraphics[width=8.5cm]{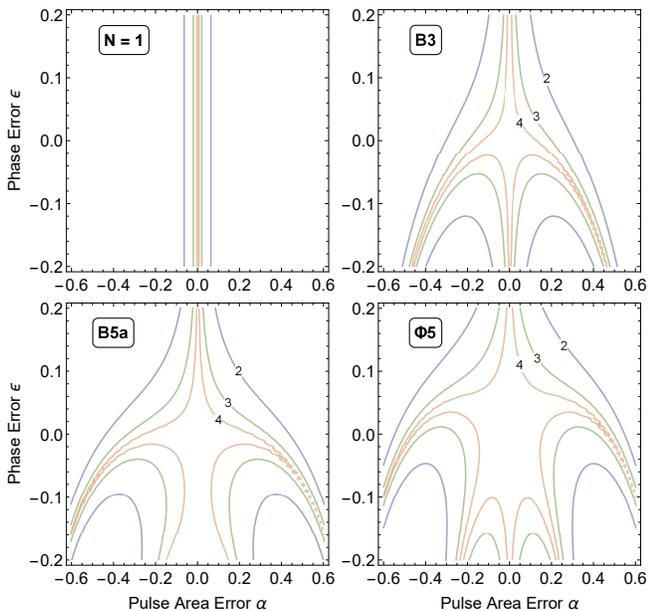}
	\caption{
Transition probability vs pulse area deviation $\alpha$ and phase error $\epsilon$ for a single pulse (top left), the B3 sequence \ref{B3} (top right), the B5a composite sequence \eqref{B5a} (bottom left) and the five-pulse phase-error compensating sequence \eqref{F5} (bottom right).
The numbers $m=2,3,4$ on the contours correspond to probability levels of $1-10^{-m}$.}
	\label{fig:5}
\end{figure}
%=================================================================

We now consider a sequence of five pulses.
Similarly to the $N=3$ case, we calculate the product in Eq.~\eqref{U^N2} and expand $U^{(5)}_{11}$ at $(\alpha=0,\epsilon=0)$.
We obtain for the first non-zero coefficients in the expansion the expressions
\bse
\begin{align}
s_{10}=&-\frac{\pi}{2}\alpha \left[ 1+2\cos(\phi_2-\phi_3)+2\cos(2\phi_2-\phi_3)\right] ,\\
s_{11}=& \pi\alpha\epsilon\left[(\phi_2-\phi_3)\sin(\phi_2-\phi_3)\right. \notag \\
&\left. \qquad +(2\phi_2-\phi_3)\sin(2\phi_2-\phi_3)\right] .
\end{align}
\ese
The set of equations $s_{10}=0$ and $s_{11}=0$ does not have an analytic solution because the latter of these is transcendental.
The values of $\phi_2$ and $\phi_3$, which nullify $s_{10}$ and $s_{11}$, can be found numerically.
One of the solutions is (approximately) $\phi_2 = 0.743\pi$, $\phi_3 = 0.395\pi$, and hence the composite sequence reads
\be\label{F5}
\Phi5 = \pi_0 \pi_{0.743\pi} \pi_{0.395\pi} \pi_{0.743\pi} \pi_0.
\ee
This composite sequence eliminates errors in the pulse area up to order $O(\alpha^2)$ and in the phases up to order $O(\epsilon^1)$.

In Fig.~\ref{fig:5} the transition probability is plotted as a function of the pulse area error $\alpha$ and the systematic error $\epsilon$ in the phases for the composite sequences B3, B5a, and  $\Phi5$.
% \be\label{B5-old}
%\pi_0 \pi_{\frac45\pi} \pi_{\frac25\pi} \pi_{\frac45\pi} \pi_0 .
%\ee
We also plot the excitation profile of a single pulse ($N=1$), which is insensitive to systematic phase error (as far as the transition probability is concerned), but it lacks compensation in the pulse area.
Obviously,  the phase-compensating composite pulse $\Phi5$ provides an ultra-accurate transition probability ($p>99.99\%$) over a much broader region than the other pulses.
Note that the previous area-compensating composite pulse B5a is fairly resistant to phase errors, although not as much as the dedicated phase-error correcting composite pulse $\Phi5$. %compared to the other B5 pulses, Eq.~\eqref{B5}
The reason is that the values of its phases are not too far from the ones of $\Phi5$, see Eqs.~\eqref{F5} and \eqref{B5a}.

\subsection{Longer composite sequences}

\begin{figure}[tb]
	\includegraphics[width=8.5cm]{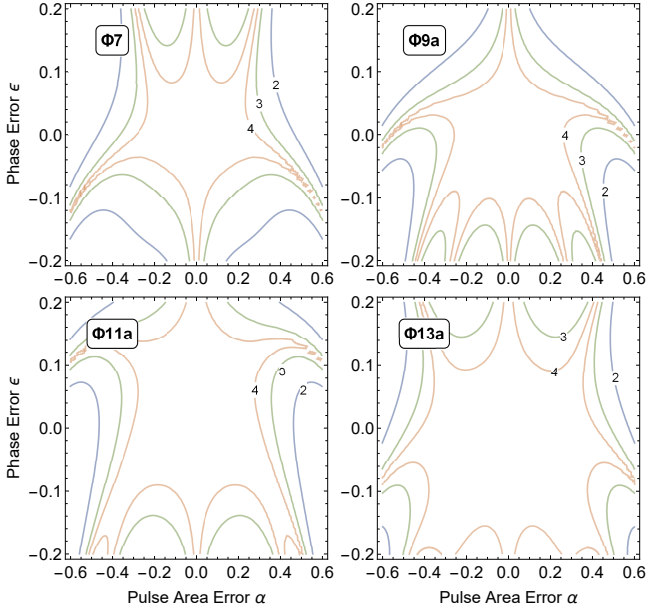}
	\caption{Transition probability vs pulse area deviation $\alpha$ and phase error $\epsilon$ for phase-error compensating sequences with $N=7,9,11,13$ pulses. The phases are given in Table \ref{tab:phases}.
The numbers $m=2,3,4$ on the contours correspond to probability level of $1-10^{-m}$.
%\textbf{VYRNI STARATA 9ka, oznachena s 9a w tablcata. Tq e nai-dobroto, koeto imame. BT: DONE}
}
	\label{fig:791113}
\end{figure}

We continue with a sequence of seven pulses, which presents us three phases to be used as free control parameters.
We choose the non-zero coefficients, which we want to nullify, to be $s_{10}$, $s_{11}$, and $s_{30}$.
The explicit expressions for these coefficients are too cumbersome to be presented here, but their numeric cancellation is straightforward.
In such a way, we obtain numerous solutions for the phases, one of which is $\phi_2=0.591\pi$, $\phi_3=-0.307\pi$, $\phi_4=-0.575\pi$.
Hence the corresponding seven-pulse composite sequence reads
\be\label{F7}
\Phi7 = \pi_0 \pi_{0.591\pi} \pi_{-0.307\pi} \pi_{-0.575\pi} \pi_{-0.307\pi} \pi_{0.591\pi} \pi_0.
\ee
Figure~\ref{fig:791113}(top left) shows the excitation profiles of this phase-compensating composite sequence.
Clearly, the high-probability area is larger than for the $\Phi5$ sequence in Fig.~\ref{fig:5}.

%=================================================================
%=================================================================

For longer sequences we can proceed in a similar manner.
The additional free phases allow us to cancel larger-order errors in the pulse area and the composite phases.
The values of all phases, which are derived, as well as the corresponding $s_{ij}$ terms, which are cancelled, are summarized in Table~\ref{tab:phases}.
In Fig.~\ref{fig:791113} we plot the excitation profiles for sequences of length $N=7,9,11,13$.
A systematic improvement of the excitation profile is observed as the length of the composite sequences increases, with the tolerance ranges exceeding 40\% for pulse area errors and 10\% for phase errors.

\begin{table}[htbp]
	\centering
	\caption{
Nullified $s_{jk}$ terms and the corresponding nominal phases for phase-error compensating composite sequences with different length $N$.
All phases are in units $\pi$.
}
	\begin{tabular}{c c}
		\toprule
		$\Phi N$ & \textbf{nullified terms and phases}  \\
		\hline
		\hline \vspace{8pt}
		%\textbf{3} & \makecell{$s_{10} $\\$(0,2/3,0)$} \\ \vspace{8pt}
		
		$\Phi 5$ & \makecell{$(s_{10},s_{11})$\\$(0,0.7433,0.3951,0.7433,0)$} \\ \vspace{8pt}
		
		$\Phi 7$ &   \makecell{$(s_{10},s_{11},s_{30})$ \\ $(0,0.5906,-0.3069,-0.5749,-0.3069,0.5906,0)$} \\ \vspace{8pt}
		
		$\Phi 9a$ & \makecell{$(s_{10},s_{11},s_{12},s_{30})$\\$(0, 0.8095, 0.5444, 1.1007, 0.1715, 1.1007, 0.5444, 0.8095, 0)$}  \\ \vspace{8pt}
	
		$\Phi 9b$ & \makecell{$(s_{10},s_{11},s_{30},s_{31})$\\$(0,1.4073,0.2688,0.6144,1.6587,0.6144,0.2688,1.4073,0)$}  \\ \vspace{8pt}
		
		$\Phi 11a$ & \makecell{$(s_{10},s_{11},s_{30},s_{31},s_{50})$\\
$(0,0.6713,0.3049,1.0965,0.7176,0.0956,$\\$0.7176,1.0965,0.3049,0.6713,0)$} \\ \vspace{8pt}

		$\Phi 11b$ & \makecell{$(s_{10},s_{11},s_{12},s_{30},s_{31})$\\
$(0, 0.6934, 0.3176, 1.1303, 0.7420, 0.1010,$ \\ $0.7420, 1.1303, 0.3176, 0.6934, 0)$} \\ \vspace{8pt}
		
		$\Phi 13a$ &  \makecell{$(s_{10},s_{11},s_{30},s_{31},s_{50},s_{51})$\\$(0, 0.8097, 0.2288, -0.0720, -0.9158, -0.1132, 0.8688,$\\$ -0.1132, -0.9158, \
			-0.0720, 0.2288, 0.8097, 0)$} \\

		$\Phi 13b$ &  \makecell{$(s_{10},s_{11},s_{12},s_{30},s_{31},s_{32})$\\
$(0, 0.8150, 0.2523, 0.6393, -0.2552, -0.4568, 0.2744)$\\$ -0.4568,-0.2552, 0.6393, 0.2523, 0.8150, 0)$} \\

		$\Phi 13c$ &  \makecell{$(s_{10},s_{11},s_{12},s_{30},s_{31},s_{50})$\\
$(0, 0.7639, 0.1842, -0.1071, -0.8840, -0.0756, 0.8432$\\$ -0.0756,-0.8840, -0.1071, 0.1842, 0.7639, 0)$} \\

		\hline
		\hline
	\end{tabular}%
	\label{tab:phases}%
\end{table}%

\subsection{Discussion}

It is important to note that in the absence of phase error ($\epsilon=0$) different composite pulses can produce the same excitation profile due to the invariance of the transition probability to various transformations of the phases, see Appendix \ref{Sec:appendix}.
However, in the presence of phase errors, the picture changes drastically.
For example, we cannot add or subtract phases $2\pi$ to/from any chosen phase because the phase error $\epsilon$ multiplies the phases $\phi_k$ and $\phi_k \pm 2\pi$ differently and hence these different phases will lead to different excitation profiles.

To this end, here we have restricted ourselves to solutions for the composite phases in the ranges $[0,2\pi]$ or $[-\pi,\pi]$.
For some of the presented composite sequences there exist other sequences of the same length which produce slightly better (i.e. broader) profiles which, however, have phases lying outside these ranges.
For example, the seven-pulse sequence with phases $(\phi _2,\phi_3,\phi_4) = (1.1703, 1.4334, 2.9010)\pi$ produces a slightly broader profile than the $\Phi 7$ sequence presented here.
We have deliberately omitted these other sequences because first, the presented sequences already produce significant phase error compensation, and second, in order to avoid ambiguity in experimental implementation.

Indeed, if the phase shifts are created by electric or magnetic fields as time-integrated Stark or Zeeman shifts then phases $\phi$ and $\phi\pm2\pi$ are physically different and it make sense to consider the respective composite sequences as different.
It is less obvious if phases $\phi$ and $\phi\pm 2\pi$ can be physically different if created by other mechanisms, e.g. by a microwave generator or an acoustooptical generator.
Therefore, in order to avoid ambiguity, we have presented only composite sequences with phases in ranges of length $2\pi$, i.e. $(0,2\pi)$ or $(-\pi,\pi)$.
It is important that any experimental realization of our sequences should consider this argument and should use the phases as reported here.

\section{Compensation of pulse area, detuning and phase errors}\label{Sec:triple}
%\sec{Error correction of several parameters}

%=================================================================
\begin{figure}[tb]
	\includegraphics[width=8.5cm]{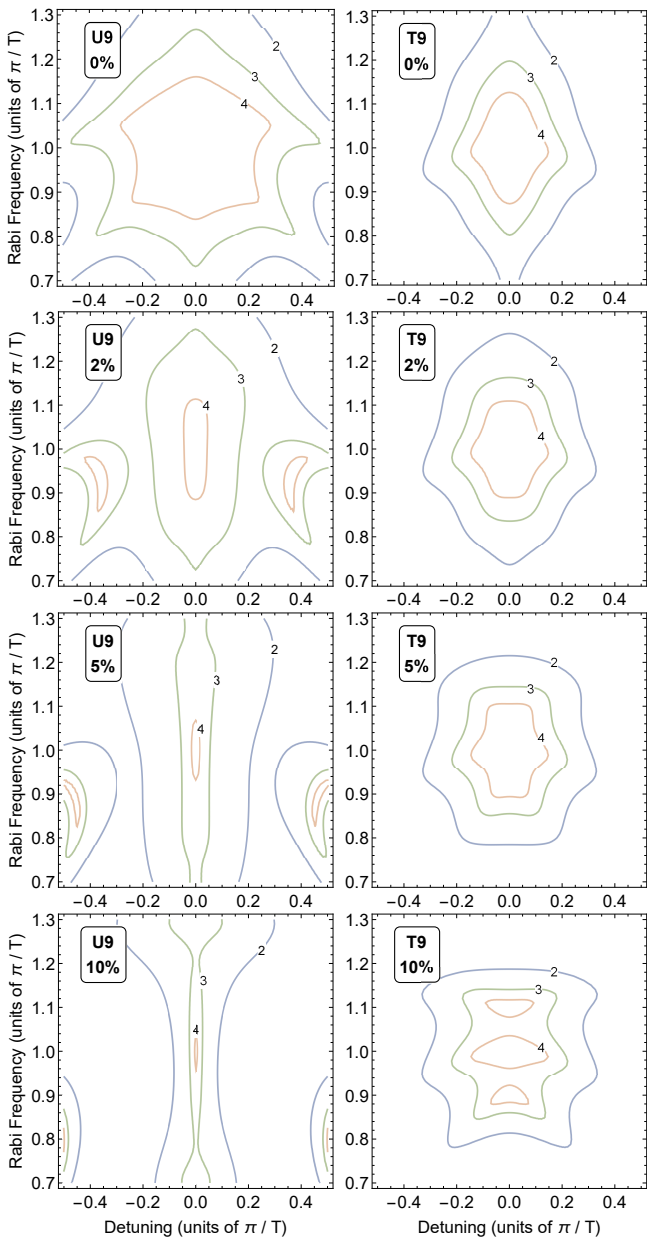}
	\caption{
Transition probability vs Rabi frequency and detuning for composite sequences of nine rectangular pulses.
The left column shows the excitation profile for the universal sequence U9 \eqref{anagram-9} with the phases of Eq.~\eqref{uni9} \cite{Genov2014PRL}, while the right column shows the profiles for the phase-error-corrected sequence T9 with the phases of Eq.~\eqref{twoDphases}.
The error in the phases is 0\%, 2\%, 5\%, and 10\%, from top frames to bottom frames.
}
	\label{TwoD}
\end{figure}
%=================================================================

We can apply the idea of phase-error compensating composite pulses to produce sequences which compensate errors in more than one parameter. 
The derivation of the phases is done in a way much similar to the one described in Sec.~\ref{Sec:method}.
For instance, in order to derive sequences, which are insensitive to errors in the Rabi frequency, the detuning, and the composite phases, we proceed as follows. First, we should specify the pulse shape in order to obtain the explicit formula for the single-pulse propagator. In our derivation, we assume rectangular pulses. Then we calculate the total propagator by taking the product of the constituent propagators. Next, we calculate the multivariate coefficients in the expansion of the total propagator vs the Rabi frequency, the detuning, and the phase error, near the point of perfect population transfer. Finally, we try to cancel as many of these derivative terms as possible.

We call these composite pulses \emph{triple compensating} and denote them with T$N$. 
For instance, the anagram composite sequence of nine rectangular pulses,
\be\label{anagram-9}
\pi_{0} \pi_{\phi_2} \pi_{\phi_3} \pi_{\phi_4} \pi_{\phi_5} \pi_{\phi_4} \pi_{\phi_3} \pi_{\phi_2}\pi_0,
\ee
with phases
%\begin{align}\label{twoDphases}
%\phi_2 = 1.348\pi, \quad \phi_3=1.257\pi, \notag\\
%\phi_4 = 0.166\pi, \quad \phi_5=0.167\pi.
%\end{align}
\be\label{twoDphases}
\phi_2 = 1.348\pi,\ \phi_3=1.257\pi,\ \phi_4 = 0.166\pi,\ \phi_5=0.167\pi.
\ee
is names as T9 and it is robust against errors in the Rabi frequency, the detuning and the composite phases.

In Fig.~\ref{TwoD} we compare the transition probabilities of this composite pulse (right column) with the transition probability, produced by a nine-pulse universal composite sequence U9, derived in Ref.~\cite{Genov2014PRL} (left column), which has the same form as Eq.~\eqref{anagram-9} but with the phases
\be\label{uni9}
\phi_2 = 0.635\pi,\ \phi_3=1.35\pi,\ \phi_4 = 0.553\pi,\ \phi_5=0.297\pi.
\ee
As seen in the figure, with attention to the 99.99\% contour (label 4), the universal sequence U9 produces a more robust excitation profile in the absence of phase errors (top frames), but in the case when phase errors are present, the phase-compensating sequence T9 beats the universal one (middle and bottom frames).

\section{Discussion and Conclusions} \label{Sec:conclusion}

In the present work we presented an approach to build composite pulses, which are insensitive to systematic errors in the composite phases.
We have shown explicit results for sequences of up to 13 pulses, with simultaneous compensation of pulse area and phase errors, but these results can be readily extended to larger number of pulses.
We have also presented triple compensation of errors in the pulse area, the frequency detuning and the phases.
These phase-error-corrected composite sequences have the potential to eliminate the main limitation of the composite pulses technique: the requirement for accurate phase control. 
They should make it possible to extend the application of the powerful and flexible concept of composite pulses to physical platforms wherein the accurate phase control is difficult or impossible.

One possible way to implement the proposed phase-compensating composite pulses is to use a recently derived generalization of the composite-pulses method to detuning pulses \cite{TorosovDetuningPulses2019}.
In that work, a sequence of \emph{detuning pulses} is used, with the areas of these pulses being the control parameters. 
In the limit of very short pulses the sequence can be seen as a composite pulse, where the composite phases are proportional to the area of the detuning pulse. 
Therefore, a systematic error naturally occurs.

Another application of these phase-error resilient composite sequences could be for achromatic devices for frequency conversion in nonlinear optics \cite{Genov2014jopt,Rangelov2014}.
There the composite approach is implemented by using nonlinear crystals of different materials and different thicknesses: alternating thick slabs of one material used as analogs of $\pi$ pulses, and thin slabs of another material used for the phase jumps (via controlled phase mismatch).
Systematic errors in the composite phases occurs naturally because the phases are proportional to the thickness of the corresponding slabs and to a scale differently for different frequencies.

\acknowledgments
%\acknowledgements
The authors acknowledge useful discussions with Andon Rangelov.
This work is supported by the European Commission's Horizon-2020 Flagship on Quantum Technologies project 820314 (MicroQC).

\appendix

\section{Non-equivalence of composite pulses in the presence of phase errors}\label{Sec:appendix}

%=================================================================
\begin{figure}[tb]
	\includegraphics[width=8.5cm]{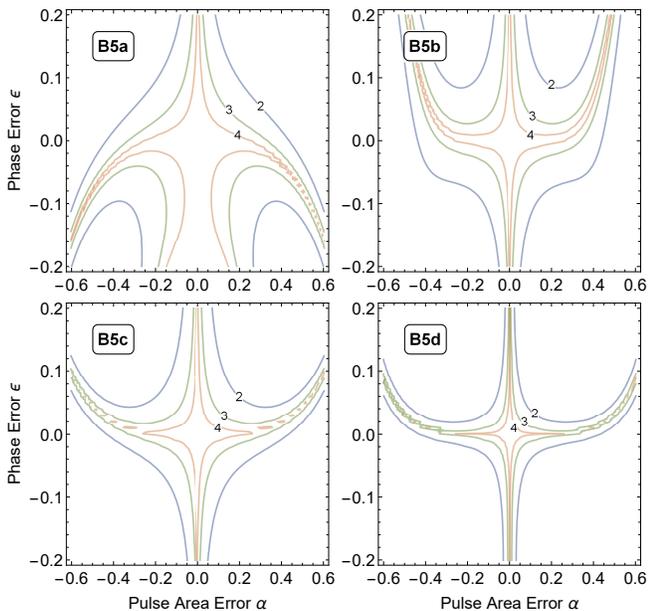}
	\caption{Transition probability vs pulse area deviation $\alpha$ and phase error $\epsilon$ for the four different B5 sequences of Eq.~\eqref{B5} (a, b, c, and d).
		The numbers $m=2,3,4$ on the contours correspond to probability level of $1-10^{-m}$.
	}
	\label{fig:diffFives}
\end{figure}
%=================================================================

In the absence of phase error ($\epsilon=0$) different composite pulses can produce the same excitation profile due to the invariance of the transition probability to various transformations of the phases \cite{Torosov2019}.
Examples of population-preserving transformations are:
 (i) the simultaneous sign flip of all composite phases, $\{-\phi_1,-\phi_2,\ldots -\phi_N\}$;
 (ii) the addition/subtraction of arbitrary integer multiples of $2\pi$ to any composite phase, $\{\phi_1+2k_1\pi,\phi_2+2k_2\pi,\ldots \phi_N+2k_N\pi\}$, where $k_j$ are arbitrary integers;
 (iii) the application of the composite sequence in the reverse order, $\{\phi_N,\phi_{N-1},\ldots \phi_1\}$;
 (iv) the addition of the same phase shift, e.g. $\phi_0$, to all phases in the sequence.
Given a composite pulse sequence, these four features allow one  to construct other composite sequences, which deliver the same transition probability.
In addition, because the composite phases are derived from a set of trigonometric equations, there are multiple solutions which cannot be obtained from each other by using the above operations but still deliver the same transition probability.

For example, the symmetric composite sequences B$n$ of Ref.~\cite{Torosov2011PRA}, which are of the type \eqref{anagram}, with phases
\bse
\be\label{Bn-1}
\phi_k = \frac{k(k-1)n}{N}\pi \quad (k=1,2,\ldots, N),
\ee
produce the same excitation profiles as the pulse sequence with the phases
\be\label{Bn-2}
\phi_k = \frac{k(k-1)}{N}\pi \quad (k=1,2,\ldots, N).
\ee
\ese

For $N=3$ pulses, both Eqs.~\eqref{Bn-1} and \eqref{Bn-2} give the well-known sequence \cite{Freeman1980,Torosov2011PRA}
\bse
\be
\text{B}3 = \pi_0 \pi_{\frac23\pi} \pi_0, \label{B3}
\ee
By inverting the sign of the second phase and adding $2\pi$ we find another equivalent sequence,
\be
 \pi_0 \pi_{\frac43\pi} \pi_0. \label{B3-43}
\ee
\ese

For $N=5$ pulses, Eqs.~\eqref{Bn-1} and \eqref{Bn-2} produce the two sequences \cite{Torosov2011PRA}
\bse\label{B5}
\begin{align}
\text{B}5a&= \pi_0 \pi_{\frac45\pi} \pi_{\frac25\pi} \pi_{\frac45\pi} \pi_0, \label{B5a} \\
\text{B}5b&= \pi_0 \pi_{\frac25\pi} \pi_{\frac65\pi} \pi_{\frac25\pi} \pi_0.\label{B5b}
\end{align}
Using the transformations described above we can generate two other sequences,
\begin{align}
\text{B}5c&= \pi_0 \pi_{\frac65\pi} \pi_{\frac85\pi} \pi_{\frac65\pi} \pi_0,\label{B5c} \\
\text{B}5d&= \pi_0 \pi_{\frac85\pi} \pi_{\frac45\pi} \pi_{\frac85\pi} \pi_0.\label{B5d}
\end{align}
\ese
All these four sequences produce the same transition probability in the absence of phase errors.
In addition we can add or subtract multiples of $2\pi$ to/from any phase in the above sequences and generate infinitely many equivalent sequences.

However, in the presence of phase errors, the picture changes drastically.
The symmetry properties (i) and (iv) listed above still stand as well as property (iii), which is irrelevant for our symmetric sequences \eqref{anagram}.
However, property (ii) is not valid any more because the phase error $\epsilon$ multiplies the phases.
Then the replacement $\phi_k\to \phi_k \pm 2\pi$ of any phase will lead to a different excitation profile.
For example, Fig.~\ref{fig:diffFives} shows the transition probability produced by the four sequences of Eqs.~\eqref{B5} versus the pulse area error $\alpha$ and the phase error $\epsilon$.
The first of these sequences, Eq.~\eqref{B5a} clearly outperforms the others in the presence of phase errors, but still underperforms the dedicated error-correcting sequence \eqref{F5}, see Fig.~\ref{fig:5}.

%%%%%%%%%%%%%%%%%%%%%%%%%%%%%%%%%%%%%%%%%%%%%%%%%%%%%%%%%%%%%%%%%%%%%%%%%%%%%%%%%%%%%%%%%%%%%%%%%
%%%%%%%%%%%%%%%%%%%%%%%%%%%%%%%%%%%%%%%%%%%%%%%%%%%%%%%%%%%%%%%%%%%%%%%%%%%%%%%%%%%%%%%%%%%%%%%%%
%%%%%%%%%%%%%%%%%%%%%%%%%%%%%%%%%%%%%%%%%%%%%%%%%%%%%%%%%%%%%%%%%%%%%%%%%%%%%%%%%%%%%%%%%%%%%%%%%
%%%%%%%%%%%%%%%%%%%%%%%%%%%%%%%%%%%%%%%%%%%%%%%%%%%%%%%%%%%%%%%%%%%%%%%%%%%%%%%%%%%%%%%%%%%%%%%%%
%%%%%%%%%%%%%%%%%%%%%%%%%%%%%%%%%%%%%%%%%%%%%%%%%%%%%%%%%%%%%%%%%%%%%%%%%%%%%%%%%%%%%%%%%%%%%%%%%

\end{document}